\documentclass[a4paper, amsfonts, amssymb, amsmath, reprint, showkeys, nofootinbib, twoside]{revtex4-1}
\usepackage[english]{babel}
\usepackage[utf8]{inputenc}
\usepackage[colorinlistoftodos, color=green!40, prependcaption]{todonotes}
\usepackage{amsthm}
\usepackage{mathtools}
\usepackage{physics}
\usepackage{xcolor}
\usepackage{graphicx}
\usepackage[left=23mm,right=13mm,top=35mm,columnsep=15pt]{geometry} 
\usepackage{adjustbox}
\usepackage{placeins}
\usepackage[T1]{fontenc}
\usepackage{lipsum}
\usepackage{csquotes}
\usepackage[pdftex, pdftitle={Article}, pdfauthor={Author}]{hyperref} 
\begin{document}
\title{Spin-orbit gaps in the $s$ and $p$ orbital bands of an artificial honeycomb lattice}

\author{J.J. \surname{van den Broeke$^\dagger$}}
\author{I. \surname{Swart$^\S$}}
\author{C. \surname{Morais Smith$^\dagger$}}
\author{D. \surname{Vanmaekelbergh$^\S$}}

\affiliation{$^\dagger$Institute for Theoretical Physics, Utrecht University, Utrecht 3584 CC, Netherlands}
\affiliation{$^\S$Debye Institute for Nanomaterials Science, Utrecht University, Utrecht 3584 CC, Netherlands}

\date{\today} 

\begin{abstract}
Muffin-tin methods have been instrumental in the design of honeycomb lattices that show, in contrast to graphene, separated $s$ and in-plane $p$ bands, a $p$ orbital Dirac cone, and a $p$ orbital flat band. Recently, such lattices have been experimentally realized using the 2D electron gas on Cu(111). A possible next avenue is the introduction of spin-orbit coupling to these systems. Intrinsic spin-orbit coupling is believed to open topological gaps, and create a topological flat band. Although Rashba coupling is straightforwardly incorporated in the muffin-tin approximation, intrinsic spin-orbit coupling has only been included either for a very specific periodic system, or only close to the Dirac point. Here, we introduce general intrinsic and Rashba spin-orbit terms in the Hamiltonian for both periodic and finite-size systems. We observe a strong band opening over the entire Brillouin zone between the $p$ orbital flat band and Dirac cone hosting a pronounced edge state, robust against the effects of Rashba spin-orbit coupling.  

\end{abstract}


\maketitle

\section{Introduction} \label{sec:outline}
	Ever since the prediction of the quantum spin Hall effect in graphene as a result of intrinsic spin-orbit coupling by Kane and Mele \cite{kane2005quantum}, efforts have been made to observe the predicted state. Not only would the intrinsic spin-orbit coupling turn graphene into a bulk insulator, it would also create conducting edge modes that would be protected from scattering by the topology of the system \cite{kane2005quantum}. Unfortunately, the gap turned out to be too small for practical applications \cite{sichau2019resonance}. Although efforts have been made to enhance the intrinsic spin-orbit coupling \cite{weeks2011engineering,kou2017two}, no convincing method has been found so far. An alternative route to study intrinsic spin-orbit effects in a honeycomb lattice is by creating an artificial lattice by placing energy barriers on top of a (heavy) metal with a 2D electronic surface gas using a scanning tunnelling microscope. This method was pionereed by Gomes et al.~\cite{gomes2012designer} using CO molecules as energy barriers on top of Cu(111). Later, Gardenier et al.~\cite{gardenier2020p} extended the method and created artificial honeycomb lattices on Cu(111) in which the $s$ orbital and $p$ orbital Dirac bands were separated, and the $p$ orbital bands included a Dirac cone and a flat band, as predicted by tight-binding methods \cite{wu2007flat, wu2008p,beugeling2015topological,cano2018topology, van2014tight, kalesaki2014dirac}.

	The system of in-plane $p$ orbital bands is of high interest to study the effects of intrinsic spin-orbit coupling. Not only is the effect of the coupling shown to be larger in these systems \cite{scammell2019tuning,beugeling2015topological,kalesaki2014dirac} because it is onsite instead of next-nearest neighbour, as would be the case for $s$ and $p_z$ orbitals, but it is also predicted to generate topological flat bands \cite{cano2018topology}. Flat bands are particularly interesting to study interactions, as the kinetic energy is quenched. Thus, interaction driven phenomena like superconductivity and charge density waves become accessible. Moreover, flat bands can be even more interesting when they are topological \cite{ma2020spin}. Thus, introducing intrinsic spin-orbit coupling in artificial electron lattices could open up exciting new possibilities for the field.

	Patterned lattices are usually described as a two-dimensional (2D) free-electron gas confined to the lattice using a modulated (muffin-tin) potential. This muffin-tin method has yielded remarkably accurate predictions for effectively spinless systems, and has proven to be a vital tool in the design of artificial electronic lattices \cite{li2016designing,park2009making,slot2017experimental,kempkes2019robust,slot2019p,gardenier2020p,franchina2020engineering}. For these systems, the muffin-tin approach is often more convenient than the tight-binding method because there are only a few parameters involved, namely the potential landscape $V(x,y)$ and the electron effective mass. These parameters only have to be determined once for a material and patterning technique, whereas the tight-binding approach requires new fitting for each design. The muffin-tin method has the additional advantage that it does not require any assumption about the orbital character of the system. It can be complemented with a tight-binding parametrization, enabling one to understand which lattice orbitals are involved in the band formation. Besides this, the muffin-tin method is also well suited for finite-size calculations, and thus very useful for the study of edge states in practical systems.  
	
	Unfortunately, the inclusion of intrinsic spin-orbit coupling in muffin-tin models is not straightforward. The theoretical approaches proposed so far vary substantially \cite{ghaemi2012designer,scammell2019tuning}, and are only valid for a specific setup or only describe the physics near the Dirac point. In addition, these techniques are not easily extended to finite-size calculations.   
	
	Here, we propose a heuristic method using text book spin-orbit Hamiltonian terms that have very few input parameters and reproduce the defining features of other approaches. Additionally, this method allows for calculations on finite-size systems, as is vital for topological applications.
	
	The outline of this paper is the following: in Sec.~\ref{sec:model}, we review the muffin-tin model and adapt it to incorporate Rashba and intrinsic spin-orbit coupling. We then investigate the influence of spin-orbit coupling on a honeycomb toy model in Sec.~\ref{sec:honeycomb}, an present our conclusions in Sec.~\ref{sec:conclusions}. 
    
\section{The model} \label{sec:model}
 Let us consider an artificial lattice, created by adatoms arranged to form an anti-lattice to the underlying electrons from the the surface state of the substrate. These can be approximated as a 2D electron gas with an effective mass $m^*$, and the system can be described by the one-electron time-independent 2D Schr\"odinger equation,
\begin{equation}\label{eq:schrodinger}
    \left(\frac{-\hbar^2}{2m^*}\nabla^2 
    +V\right) \Psi=E\Psi,
\end{equation}
where $V$ is the potential created by the adatoms patterning the surface. Thus, the only freedom in the input parameters is in the shape of the potential. When modeling a patterned potential $V(x,y)$ as a collection of disk shaped protrusions, also called a muffin-tin potential, only two parameters remain, namely the disk height and the disk width. 

\begin{figure}
\centering
\includegraphics[scale=0.25]{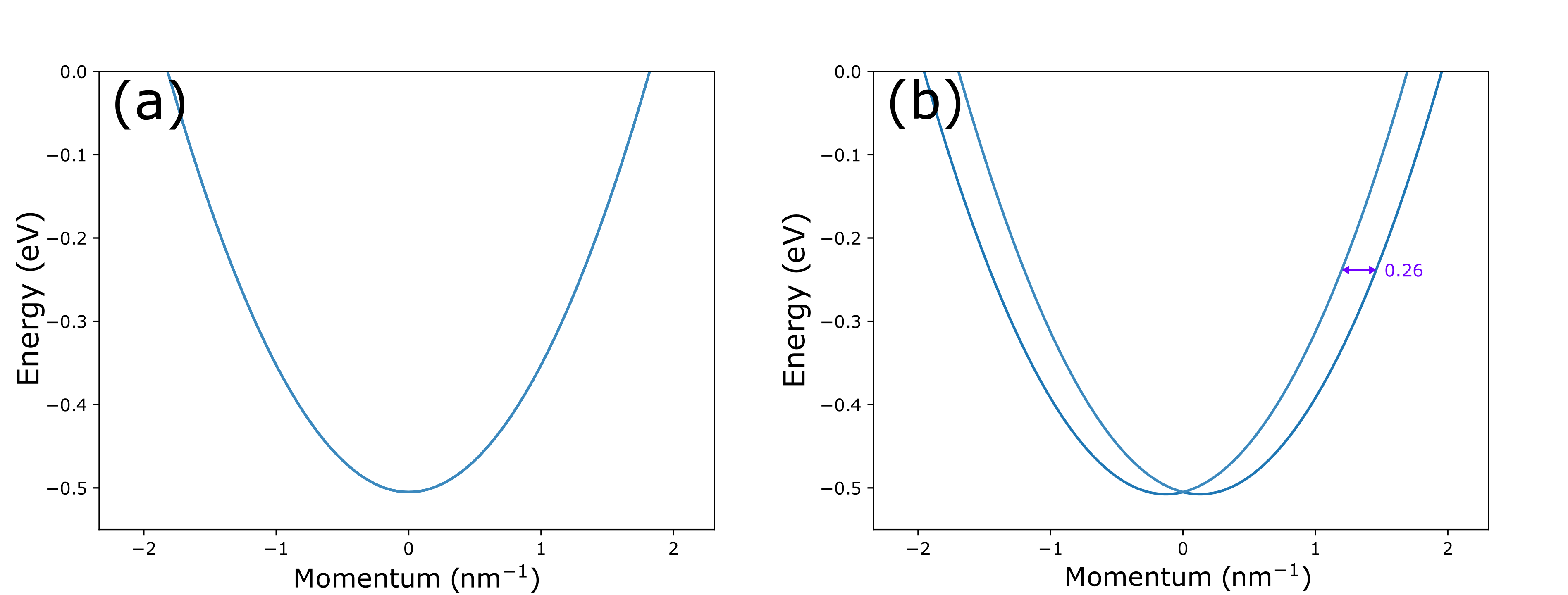}
\caption{The gold surface state, calculated as a free electron gas (a) without and (b) with Rashba spin-orbit coupling. Here, an effective mass of 0.25 em and  $\alpha_1=6.02 \times 10^4$ m/ s were used, as measured in Ref.~\cite{hoesch2004spin}}
\label{fig:rashba}
\end{figure}

In order to study the effect of spin-orbit coupling in artificial lattices, we start with the spin-orbit coupling that originates from the Dirac equation as a relativistic correction to the Schrodinger equation, given by
\begin{equation}\label{eq:spinorbit}
    H_{SO}=\frac{\hbar}{4 m^2 c^2}(\nabla V^* \times \bold{p}) \cdot \bold{\sigma},
\end{equation}
where $m$ is the real electron mass, $V^*$ is the full potential, $\bold{p}$ is the vector momentum, $\bold{\sigma}$ is the vector of Pauli matrices, and $c$ is the speed of light. Here, we see that spin-orbit coupling is proportional to the gradient of the potential $V^*$. The Rashba spin-orbit coupling originates from Eq.~(\ref{eq:spinorbit}) at the surface of materials, due to the large change in the potential at the interface. As inversion symmetry is not present, spin degeneracy is not required and is indeed broken. We can obtain the Rashba term for the muffin-tin model by considering a 2D material with a potential change in the out-of-plane direction:
\begin{equation}\label{eq:rashba}
    H_R=\alpha_1 \left(p_x \sigma_y -p_y\sigma_x \right).
\end{equation}
 Here, $p_i$ $(i=x,y)$ is the momentum component, $\sigma_i$ are the Pauli matrices, and $\alpha_1$ is the effective strength of the Rashba coupling. As a check, we add this term to Eq.~(\ref{eq:schrodinger}), and use $\alpha_1$ and $m^*$ as experimentally measured in Ref.~\cite{hoesch2004spin} for the Au(111) surface state, which is known to have a large Rashba splitting. It can be seen that our calculation reproduces the previously observed Rashba splitting of 0.26 nm$^{-1}$ between the two parabola minima \cite{hoesch2004spin}, as shown in Fig.~\ref{fig:rashba}. Therefore, the muffin-tin method can accurately describe the Rashba spin-orbit coupling. 

Next, we consider intrinsic spin-orbit coupling, which is a consequence of the coupling between the magnetic moment of the orbital angular momentum and the spin of the electron. In atomic systems, the intrinsic spin-orbit term $H_I$ scales supralinear with the atomic number \cite{atkins2018physical}. Thus, $H_I$ tends to be much larger for heavier elements. In the case of the muffin-tin technique however, the substrate is approximated as a 2D electron gas with an effective mass $m^*$ and a scattering potential originating from the patterned adatoms. Thus, details on the precise potential landscape, like the size of the nuclei in the substrate, that give rise to the intrinsic spin-orbit coupling, are lost. Here, we propose a heuristic solution to this issue that maps the intrinsic spin-orbit coupling coming from Eq.~(\ref{eq:spinorbit}) to the muffin-tin calculations by assuming an effective coupling  $\alpha_2$ between the patterned muffin-tin potential and the spin-orbit term, 
\begin{equation}\label{eq:inspinorbit1}
    H_I=\alpha_2 (\nabla V \times \bold{p}) \cdot \bold{\sigma}.
\end{equation}
By allowing this effective parameter $\alpha_2$ to be system dependent, the method can be fitted to any substrate material. Additionally, because of the relative simplicity of this approach, it easily translates to both finite and periodic calculations, which is highly convenient when working with topological materials. Please note that $\alpha_1$ and $\alpha_2$ do not have the same units and $H_R\approx H_I$ does not mean that $\alpha_1 \approx \alpha_2$, as the potential derivative is only absorbed in $\alpha_1$, cf. Eq.~(\ref{eq:rashba}) and (\ref{eq:inspinorbit1}).

As the electrons are confined to the $x,y$ plane, only the $z$ component of the cross product survives, and the intrinsic spin-orbit contribution becomes,
\begin{equation}\label{eq:inspinorbit2}
    H_I=\alpha_2\left(\frac{\partial V}{\partial x} p_y - \frac{\partial V}{\partial y} p_x\right) \sigma_z.
\end{equation}

It is hoped that with appropriate fitting for the effective parameter $\alpha _2$, Eq.~(\ref{eq:inspinorbit2}) will yield adequate predictions for spin-orbit coupling in an artificial lattice. Indeed, as shown in the next section, Eq.~(\ref{eq:inspinorbit2}) reproduces the main features found using other methods. Adding $H_R$ and $H_I$ to Eq.~(\ref{eq:schrodinger}), the full time-independent one-electron Schr\"odinger equation becomes:
\begin{multline}\label{eq:spinorbitfull}
    \left[\frac{-\hbar^2}{2m}\nabla^2 
   -i\hbar\alpha_2\left(\frac{\partial V}{\partial x} 
   \frac{\partial}{\partial y} - \frac{\partial 
   V}{\partial y} \frac{\partial }{\partial x}\right) \sigma_z \right.\\
      \left.-i\hbar \alpha_1\left(
   \frac{\partial}{\partial x}\sigma_y - 
   \frac{\partial }{\partial y}\sigma_x\right) 
       +V\right] \Psi_\sigma=E\Psi_\sigma.
\end{multline} 
For finite-size systems, solving this equation is not much different from solving the spinless system. Nevertheless, there is one important point to consider. Due to the presence of a derivative of the potential, the precise shape of the potential becomes important. For the muffin-tin potential, we would encounter infinities in the spin-orbit term. We solve this by using Gaussian potentials instead. As shown in Appendix~\ref{ap:Gaus}, a change in potential shape from muffin-tin to Gaussian does not yield significantly different results in the case without spin-orbit coupling, and is therefore an appropriate approximation.  

In the case of a periodic system, careful Fourier transformation is required to incorporate the spin-orbit couplings. We first Fourier transform the wave function:
\begin{equation}
\Psi_\sigma (\bold{x})=\frac{1}{\sqrt{A}}\sum_{\bold{k}}e^{i\bold{k}\cdot \bold{x}}\Psi_\sigma(\bold{k}),
\end{equation}
where $A=L^2$ is the system size in which the wave function is periodic, and $\bold{k}=\frac{2\pi}{L}\left(l_x,l_y\right)$, with $l_i$ ranging from $-\infty$ to $\infty$. Meanwhile, we also Fourier transform the potential $V$. However, as $V$ has the unit cell periodicity we have
\begin{equation}
    V(\bold{x})=\sum_{\bold{K}}e^{i\bold{K}\cdot \bold{x}}V_{\bold{K}},
\end{equation}
where $\bold{K}$ are the reciprocal lattice vectors. Applying these transformations to Eq.~(\ref{eq:spinorbitfull}), the resulting equation also has to hold for a single Fourier component $\bold{q}$,
\begin{multline}
     \frac{\hbar^2}{2m}q^2\Psi_\sigma(\bold{q})-\hbar\alpha_1\left(q_y\sigma_x-q_x\sigma_y\right)\Psi_\sigma(\bold{q})\\
    -i\hbar \alpha_2\sum_{\bold{K}} V_{-\bold{K}} \left(K_x q_y-  K_y q_x\right) \sigma_z \Psi_\sigma(\bold{q}+\bold{K})\\
    +\sum_{\bold{K}}V_{-\bold{K}}\Psi_\sigma(\bold{q}+\bold{K})
    =E_q\Psi_\sigma(\bold{q}).
\end{multline}
Only wave functions of the shape $\Psi(\bold{q})$ and  $\Psi(\bold{q}+\bold{K})$ appear in this equation. We can therefore apply a shift $\bold{q} \rightarrow \bold{q}+\bold{K}'$ and  $\bold{K} \rightarrow \bold{K}-\bold{K}'$ to obtain a coupled system of equations for each $\bold{q}$ in the Brillouin zone,
\begin{multline} \label{eq:systemSO}
    \frac{\hbar^2}{2m}(\bold{q}+\bold{K}')^2\Psi_\sigma(\bold{q}+\bold{K}')\\
    -\hbar\alpha_1\left[(q_y+K_y')\sigma_x-(q_x+K_x')\sigma_y\right]\Psi_\sigma(\bold{q}+\bold{K}')\\
    -i\hbar \alpha_2\sum_{\bold{K}} V_{\bold{K}'-\bold{K}} \left[(K_x-K_x') q_y-  (K_y-K_y') q_x+ \right. \\
   \left. K_x K_y' - K_y K_x'\right] \sigma_z \Psi_\sigma(\bold{q}+\bold{K})\\
    +\sum_{\bold{K}}V_{\bold{K}'-\bold{K}}\Psi_\sigma(\bold{q}+\bold{K})
    =E_{\bold{q}+\bold{K'}}\Psi_\sigma(\bold{q}+\bold{K}').
\end{multline}
 In principle, Eq.~\ref{eq:systemSO} is an infinite set of equations, one for each $\bold{K}'$, and can therefore not be solved. However, we are only interested in the lowest bands. As $V_{\bold{K}}$ becomes exponentially small for large values of $\bold{K^2}$ for both Gaussian and muffin-tin potentials (see also Appendix \ref{ap:Gaus}), we can introduce a cutoff in the values of $\bold{K}'$ that we consider. We can then solve the system of equations for arbitrary $\bold{q}$ in the Brillouin zone. In this work, a square grid $i K_1+j K_2$ with $K_{1,2}$ the reciprocal primitive vectors and $i$ and $j$ integers ranging from -4 to 4, was used.

\section{Honeycomb structures}\label{sec:honeycomb} 
\begin{figure}
\centering
\includegraphics[scale=0.55]{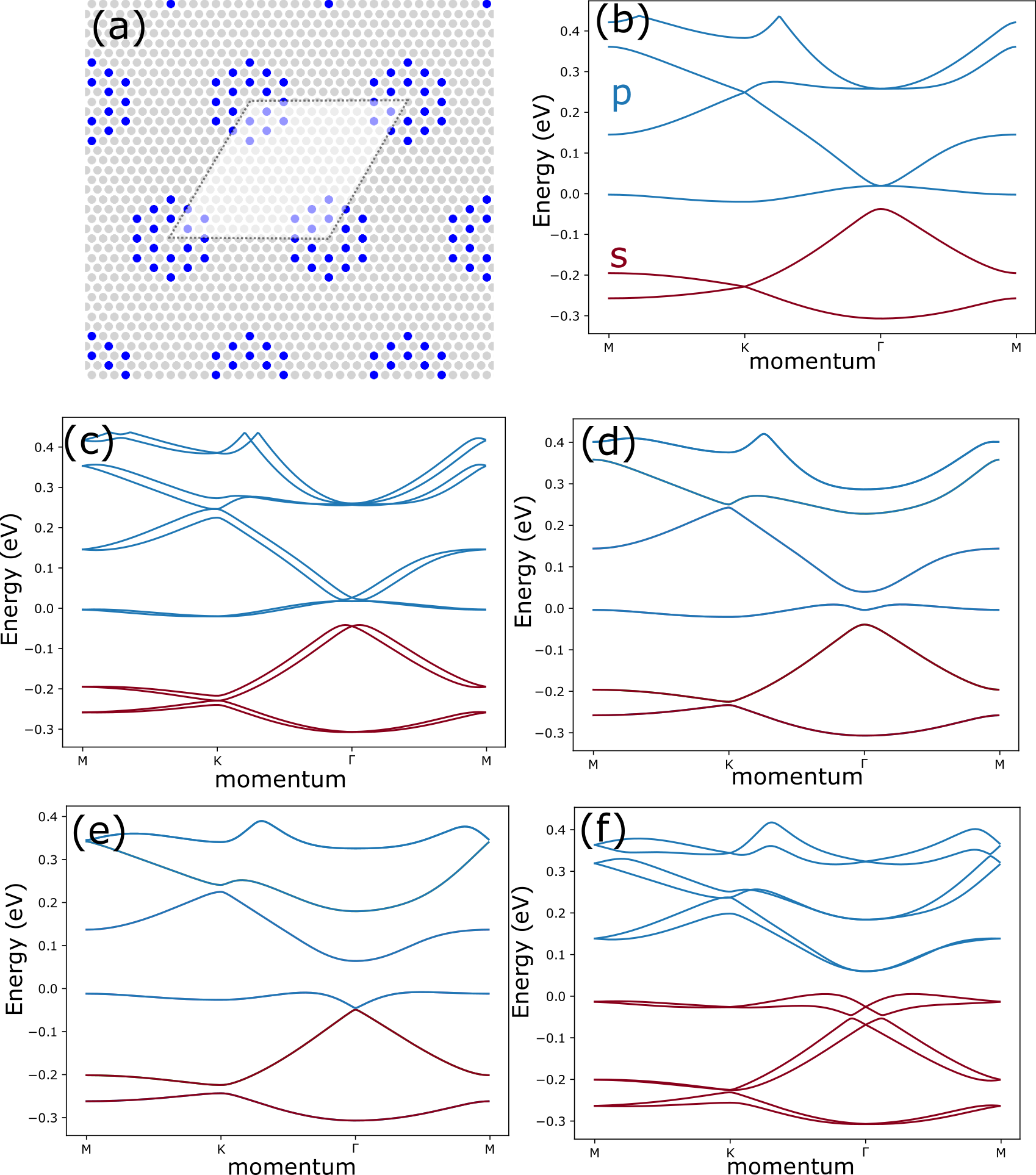}
\caption{Periodic system calculations, using Cu (111) parameters as a test case (effective mass $m^*$=0.42 em, CO molecules as Gaussians with a height of 0.45eV and a FWHM of 0.6 nm). (a) shows the arrangement of CO molecules on Cu (111) in the reference system realised experimentally in Ref.~\cite{gardenier2020p} without spin-orbit coupling. The band structure with $s$ and $p$ orbital bands plotted in red and blue respectively are shown (b) without any spin-orbit coupling; (c) with only Rashba spin-orbit coupling ($\alpha_1=1.6\times10^{4}$m/s); (d,e) with only intrinsic spin-orbit coupling  ($\alpha_2=0.8\times10^{15}$s/kg, $\alpha_2=2\times10^{15}$s/kg) and (f) with both Rashba and intrinsic spin-orbit coupling ($\alpha_1=1.6\times10^{4}$m/s, $\alpha_2=2\times10^{15}$s/kg). }
\label{fig:refsystem}
\end{figure}

In order to see the effect of the spin-orbit terms introduced above on the band structure of artificial lattices, it is instructive to first investigate a test system. For this, we will consider a honeycomb lattice, as spin-orbit coupling has been extensively studied in graphene-like lattices through other methods \cite{kane2005quantum, weeks2011engineering,cano2018topology}. As a starting point, the first artificial graphene lattice realized by Gomes et al.~\cite{gomes2012designer} using CO on the copper (111) surface might appear as a good choice. However, more elaborated designs of honeycomb lattices have recently been shown to lead to interesting features, like the appearance of a flat $p$ band \cite{gardenier2020p}. Additionally, for patterned quantum wells, intrinsic spin-orbit coupling is predicted to open a larger band gap between these higher bands than at the Dirac cone between the lower (i.e. predominantly $s$ orbital) energy bands \cite{scammell2019tuning}. We therefore use the system described in Ref.~\cite{gardenier2020p} as a reference system. This system also uses CO molecules on copper (111), but instead of positioning single CO molecules in a triangular lattice as in Ref.~\cite{gomes2012designer}, clusters of CO molecules are used. The clusters consist of two highly symmetrical rings. This added structure gives more confinement to the surface electrons without breaking the symmetry, which leads to a clear separation of $s$ and $p$ orbitals and the appearance of not only $s$ bands, as in Ref.~\cite{gomes2012designer}, but also (nearly flat) $p$ bands and a $p$ character Dirac cone. The cluster arrangement of the CO molecules on the Cu (111) surface is shown in Fig.~\ref{fig:refsystem} (a). In Ref.~\cite{gardenier2020p}, a muffin-tin potential with a height of 0.9 eV and a diameter of 0.6 nm is used. When switching to Gaussians, the choice was made for Gaussians with a full width at half maximum of 0.6 nm. With an adjustment of the potential height to 0.45 eV, this setup fully reproduces the muffin-tin results from Ref.~\cite{gardenier2020p}, as shown in Appendix~\ref{ap:Gaus}. 

The bare band structure of the reference system is shown in Fig~\ref{fig:refsystem} (b). Here, we see the two lowest energy bands forming the well known Dirac cone at the K point, like in graphene. The $p$ bands start with a (nearly) flat band, connected to two bands forming a $p$ orbital Dirac cone at the K point, which is connected to a fourth $p$ orbital band. Upon inclusion of the Rashba coupling, spin degeneracy is lifted everywhere except at the $\Gamma$ point, as shown in Fig.~\ref{fig:refsystem} (c). This result is analogous to that of previous tight-binding studies on single-orbital honeycomb lattices \cite{van2010rashba,zarea2009rashba}. Additionally, the splitting of Dirac cones under the influence of Rashba spin-orbit coupling \cite{van2010rashba} is recovered, as shown in Appendix \ref{ap:Dirac}. If only the intrinsic spin-orbit coupling is included, as shown in Fig.~\ref{fig:refsystem} (d,e), the spin degeneracy remains, and instead we see gaps opening up between the original band touching points. Indeed, this is also the result of intrinsic spin-orbit coupling in numerous other theoretical studies on honeycomb systems \cite{beugeling2015topological,scammell2019tuning,reis2017bismuthene,ghaemi2012designer, cano2018topology, van2010rashba,zarea2009rashba}. There is both theoretical and experimental evidence for these gaps to be topological and harbor protected edge states \cite{beugeling2015topological,scammell2019tuning,reis2017bismuthene,ghaemi2012designer, cano2018topology,kalesaki2014dirac}. Notably, the gap opening up between the first two $p$ orbital bands at the $\Gamma$ point is much larger than the gaps opening up at the $K$ points. In previous works, a similar trend of larger gaps between the $p$ orbital bands than between the $s$ orbital ones is observed as a consequence of the same intrinsic spin-orbit coupling \cite{beugeling2015topological,scammell2019tuning,reis2017bismuthene}. This effect can be explained by the angular momentum of $p$ orbitals, making intrinsic spin-orbit coupling an onsite effect. In $s$ orbitals that have no angular momentum, the spin-orbit coupling can only emerge through next-nearest-neighbour coupling, which connect the same sublattice in a honeycomb geometry. On the other hand, for $p$ orbitals the intrinsic spin-orbit coupling can couple $p_x$ and $p_y$ orbitals on the same site, thus rendering the effect more robust \cite{beugeling2015topological}. Additionally, we see an unexpected effect, namely, if $\alpha_2$ is large enough, the $p$ orbital flat band is no longer isolated as in Fig.~\ref{fig:refsystem} (d), but the gap between the $s$ and $p$ type bands seemingly closes to form a Dirac cone at the $\Gamma$ point, as shown in Fig.~\ref{fig:refsystem} (e). However, a small gap can be observed between the two bands. A zoom in on the gap can be found in Appendix \ref{ap:Dirac2}. This phenomenon is interesting and it has not been observed before, as far as the authors are aware. Finally, we can also include both Rashba and intrinsic spin-orbit coupling, as shown in Fig.~\ref{fig:refsystem} (d). Here, we see that the Rashba coupling can close the gaps opened by the intrinsic spin-orbit coupling, diminishing the protection of possible topological states. However, the gap between the first two $p$ orbitals is remarkably robust to the Rashba coupling. This robustness against Rashba spin-orbit coupling is of high importance in applications of topological materials, as Rashba spin orbit coupling is to a certain degree always present in devices based on 2D materials.

\begin{figure*}
\centering
\includegraphics[scale=0.6]{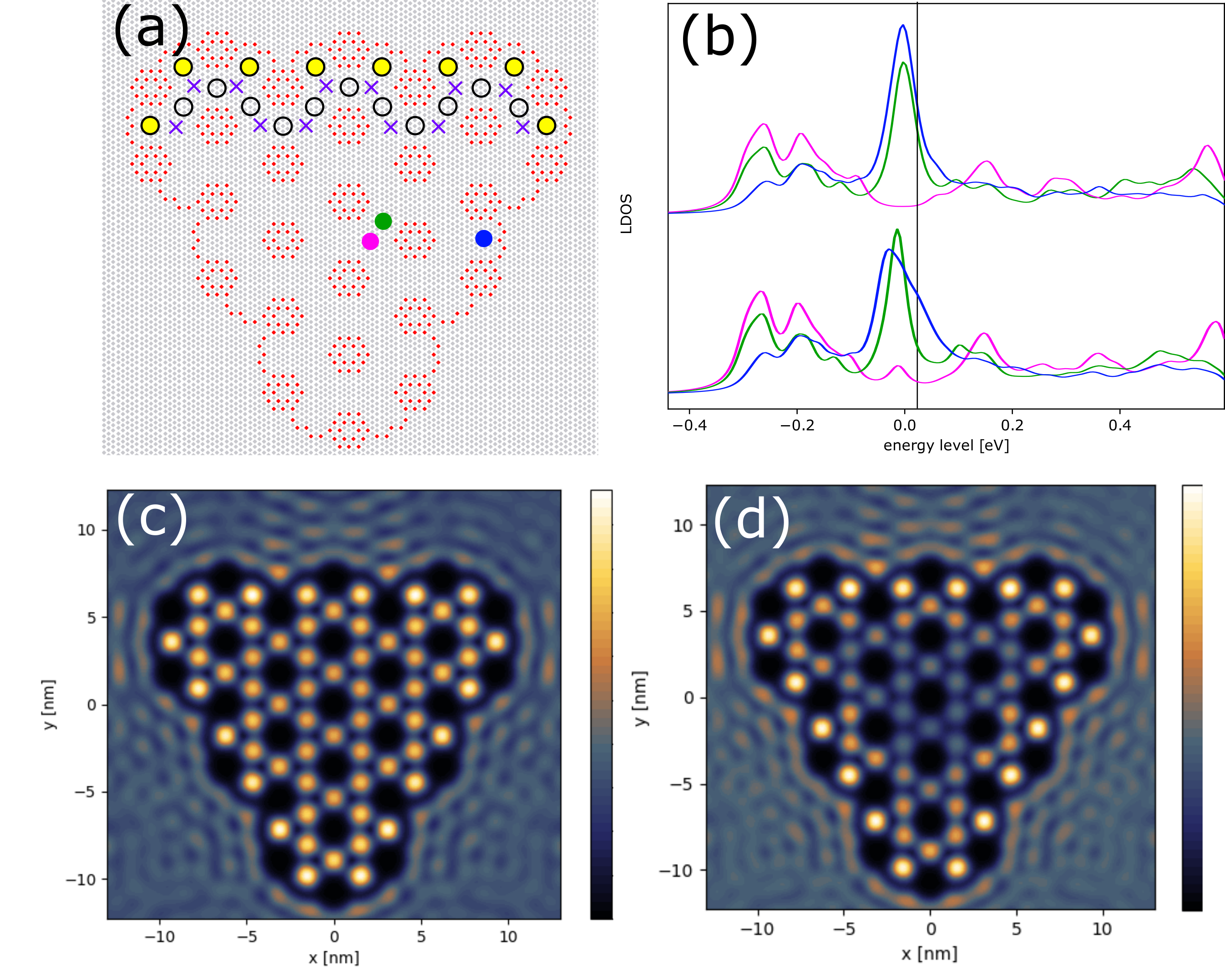}
\caption{Finite-size system calculations using Cu (111) parameters as a test case as in Fig.~\ref{fig:refsystem}. Here, a spectral broadening of 40 meV realistic for the CO on Cu (111) system has also been included. (a) The locations of CO molecules (red) on the Cu (111) grid (gray). Along the top edge, the locations of lattice sites, edge bridge sites and sub edge bridge sites have been indicated with purple crosses, yellow disks and open disks respectively. (b) The spectra calculated for the spots indicated in (a). The top and bottom curves correspond to the system without and with intrinsic spin-orbit coupling ($\alpha_2=2\times10^{15}$s/kg), respectively. (c), (d) A map of the calculated LDOS on the energy indicated by a vertical line in (b) at $E=0.027$ eV without and with intrinsic spin-orbit coupling , respectively.}
\label{fig:finiterefsystem}
\end{figure*}

In order to further study the topological nature of band gaps opened by the intrinsic spin-orbit coupling, we calculated the local density of states (LDOS) of the finite lattice. This is very helpful, as it allows to detect the existence of in-gap edge localized states, as shown in Fig.~\ref{fig:finiterefsystem}. The design used for the finite-size system is shown in Fig.~\ref{fig:finiterefsystem} (a). We study two types of locations here, onsite locations indicated by the pink dot, and bridge locations, indicated by the  green dot for the bulk and blue dot for the edge. Along the top edge, onsite edge locations have been marked with purple crosses, and  edge and sub-edge bridge sites have been marked with yellow and open circles, respectively. This design is different from the periodic case only at the boundary, where blocker potentials have been placed to separate the lattice from the surrounding 2D electron gas. The introduction of these blockers is crucial as without them there would be no clear boundary and therefore it would not be possible to study edge states. The location of blocking potentials is non trivial, as the introduction of out of lattice potentials can change the onsite energy of nearby sites. They have, therefore, been chosen in such a way as to not shift the LDOS spectra at the edge sites with respect to the bulk, as can be seen by comparing the blue and green lines in the top graph of Fig.~\ref{fig:finiterefsystem} (b). A triangular design was chosen to optimise the distance between the boundaries and at the same time have edges as uniform as possible, given the small system size. The system was studied without and with intrinsic spin-orbit coupling. The LDOS spectra of the two systems are mostly similar in the bulk, as shown in Fig.~\ref{fig:finiterefsystem} (b). The pink and green spectra representing the bulk both with and without spin-orbit coupling display two peaks from the $s$ orbitals at $\approx$-0.3 and $\approx$-0.2 eV, and both show a peak corresponding to the flat $p$ band around 0 eV. These peak locations are inline with the band structure in Fig.~\ref{fig:refsystem} (b) and (e). However, some differences are visible. In the case without spin-orbit coupling (Fig.~\ref{fig:finiterefsystem} (b) top), there is very little mixing between the $s$ and $p$ orbitals resulting in an on-site (pink) dip at the flat band energy due to negative interference between the $p$ orbitals in the flat band. In the intrinsic spin-orbit case, there is more mixing, as evidenced by the band touching between the highest $s$ and lowest $p$ orbital in Fig.~\ref{fig:refsystem} (e). Therefore, there is some onsite density of state in this case, as evidenced by a pink peak around 0 eV (Fig.~\ref{fig:finiterefsystem} (b) bottom). At a slightly higher energy (0-0.06 eV) we see a dip in the bulk spectra of the intrinsic spin-orbit case. This corresponds to the band gap between the first two $p$ orbitals in Fig.~\ref{fig:refsystem} (e) from 0 till 0.05 eV. Notably, the edge bridge site where the spectrum was calculated (blue) has an increased LDOS in this range (the shoulder is a signature of an additional peak, which cannot be separated from the flat band one), signaling a possible edge state.  Indeed, if we look at LDOS maps at 0.027 eV, we see a state clearly localized on the (sub) edge bridge sites of the system indicated in Fig.~\ref{fig:finiterefsystem} (a) for the intrinsic spin-orbit system, as shown in Fig.~\ref{fig:finiterefsystem} (d), whereas none is present for the system shown in Fig.~\ref{fig:finiterefsystem} (c), without  spin-orbit coupling. Thus, as expected, intrinsic spin-orbit coupling results in a strong decrease in the DOS over the entire bulk of the finite system in the gap between the third and fourth band, accompanied with an increase in the intensity at the bridge sites at the system edge, pointing to a protected edge state. The intensity is maximal on the bridge sites, in accordance with the $p$ orbital bands. We cannot test the precise topological nature of the edge state within the present muffin tin model; but an atomistic tight-binding study on a semiconductor system showed that this state is a helical quantum spin Hall state \cite{cano2018topology}.

\section{Conclusion} \label{sec:conclusions}
We have presented an effective muffin-tin model by introducing the intrinsic and Rashba spin-orbit coupling into the Schrödinger equation and tested the adequacy of the model by comparison with established results. Then, we have studied the effect of spin-orbit coupling on a honeycomb system with separated $s$  and $p$ orbital bands, which allows us to study $p$ orbital physics in the honeycomb system. Besides the expected band openings at the Dirac points, intrinsic spin-orbit coupling shifts the $p$ orbital flat band downwards, causing hybridization with the $s$ bands. As a result, a broad gap arises between the third and fourth band of the system; our results on a finite lattice show the emergence of an edge state in this gap. We see that Rashba spin-orbit coupling reduces the spin-orbit gaps at the K and K’ points of the Brillouin zone, but the broad gap between the third and fourth band remains robust. We should also remark that the model that we developed can be used to study the effects of spin-orbit coupling on any type of artificial lattice. 

Experimentally, strong spin-orbit coupling might be realized in artificial lattices by using a metallic surface state on a heavy element metal such as rhenium, lead or bismuth, and/or using heavy adatoms as potential barriers or attractive sites for the surface electrons. A similar concept has been reported for graphene, by placing In and Tl atoms in the hollow sites of a graphene monolayer \cite{weeks2011engineering}. Real devices, applicable in electronics, can be achieved by nanoscale patterning of heavy-element semiconductor quantum wells, such as Ge, GaAs and InSb, with a honeycomb or another geometry of interest \cite{franchina2020engineering,wang2018observation,post2019triangular}.

\section*{Acknowledgements} \label{sec:acknowledgements}
    Financial support from the European Research Council (Horizon 2020 “FIRSTSTEP”, 692691 and "FRACTAL", 865570  ) is gratefully acknowledged. We would like to thank S. E. Freeney for help with visualising the design, and  S. J. M. Zevenhuizen and R. Ligthart for help in developing an earlier non spin resolved version of the muffin-tin code.

\bibliography{SO}

\appendix

\section{Gaussian potentials\label{ap:Gaus}}
Recent works have been modeling the experimental results of artificial lattices built using CO on copper (111) by using a Muffin-tin calculation. Here, the Schr\"odinger equation is solved in two dimensions for a potential landscape where the CO molecules are modeled as positive disk shaped protrusions. This approach is convenient in Fourier space, as the Fourier transformed form of this potential is analytically known.

\begin{figure}
\centering
\includegraphics[scale=0.3]{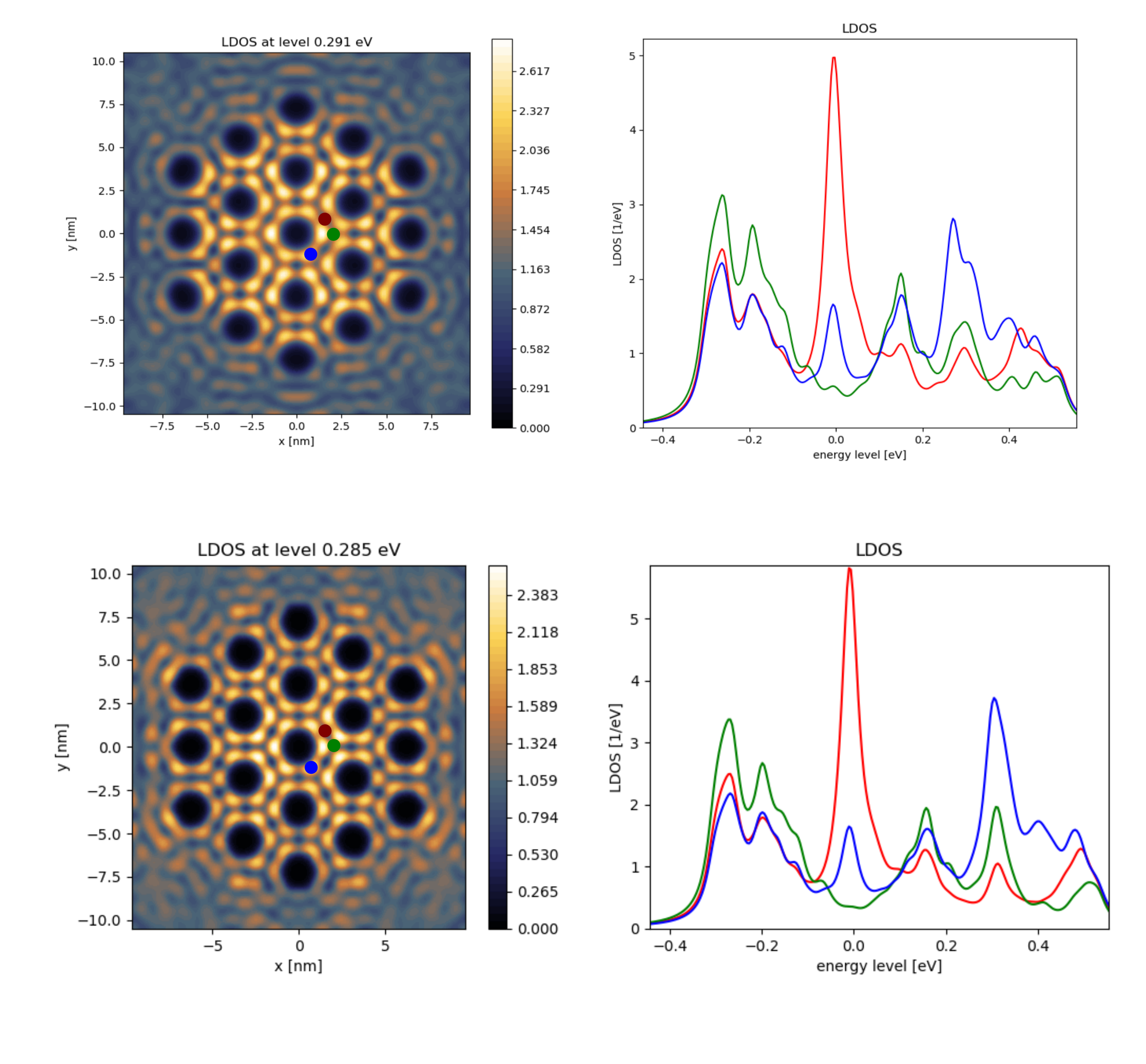}
\caption{Finite-size system comparison. Top images are made using Gaussians with a full width at half maximum (FWHM) of 0.6 nm and a heigth of 0.45eV and the bottom images were created using a classical muffin-tin calculation with a diameter of 0.6 nm and a heigth of 0.9eV. }
\label{fig:comparisson}
\end{figure}

\begin{figure}
\centering
\includegraphics[scale=0.6]{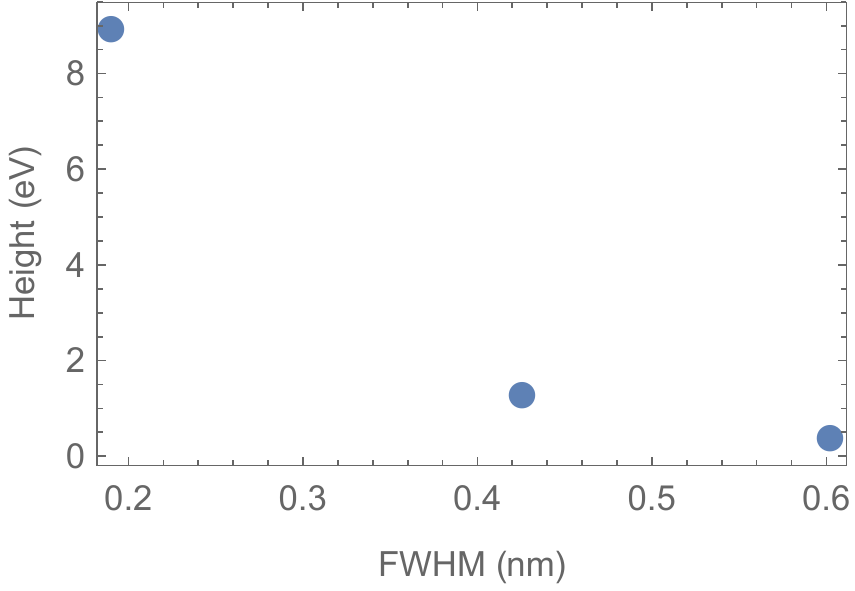}
\caption{Relation between the height of the Gaussians and their full width at half maximum (FWHM) that reproduce the muffin-tin results.}
\label{fig:relation}
\end{figure}

However, the intrinsic spin-orbit coupling term contains a derivative with respect to the potential, making the muffin-tin approach less ideal. We therefore turn to modeling the CO molecules as Gaussian potential barriers. We find that this reproduces the muffin-tin results. Furthermore, there appears to be a lot of freedom in choosing the  width of the Gaussians, as long as the height is adjusted as well (the broader the Gaussian, the lower it should be). The classical muffin-tin results, along with the results for a Gaussian potential landscape, are shown in Fig.~\ref{fig:comparisson}. The relation between the width of the Gaussian and its height in order to reproduce the classical muffin-tin results is shown in Fig. \ref{fig:relation}.

In case of a periodic system, we run into another important point. Gaussian potentials are only periodic by approximation. It is therefore not possible to analytically calculate the Fourier transform. However, as the standard deviation of the Gaussians used is several times smaller than the lattice vector, we can approximate the Fourier transform of the Gaussian in the unit cell as an infinite Fourier transform. The error is minimal, as the Gaussian potential exponentially decreases away from the center and is thus very small outside of the unit cell. We have
\begin{multline}
   V_{\bold{K}}=
   \frac{1}{A} 
   \int_{\rm{unit cell}} e^{-a |\bold{x}|^2} e^{i\bold{K}\cdot \bold{x}}d\bold{x}\\
   \approx
   \frac{1}{A}
   \int_{-\infty}^\infty e^{-a |\bold{x}|^2} e^{i\bold{K}\cdot \bold{x}}d\bold{x},
\end{multline}
where $A$ is the unit cell surface. The integral on the right is a known integral, and thus we get
\begin{equation}
    V_{\bold{K}}\approx\frac{\pi}{a} e^{-\left|\bold{K}\right|^2/4a}.
\end{equation}
Indeed, the potential becomes exponentially small for large $\bold{K}$, thus making it possible to introduce a cutoff on the $\bold{K}$ values included in Eq. \ref{eq:systemSO} to calculate the band structure.  

In classical mufin-tin calculations, the potential is fully periodic and the Fourier transformed potential is analytically known,
\begin{equation}
    V_{\bold{K}}=\frac{\pi d}{A \left|\bold{K}\right|}J_1\left(\left|\bold{K}\right|\frac{d}{2}\right)V_0.
\end{equation}
Here, $d$ is the diameter of the muffin-tin potential disks, $V_0$ is the height of the potentials, and $J_1$ is the Bessel function. Just as for the Gaussian potential, $V_{\bold{K}}$ becomes small for large $\bold{K}$.

\begin{figure}
\centering
\includegraphics[scale=0.52]{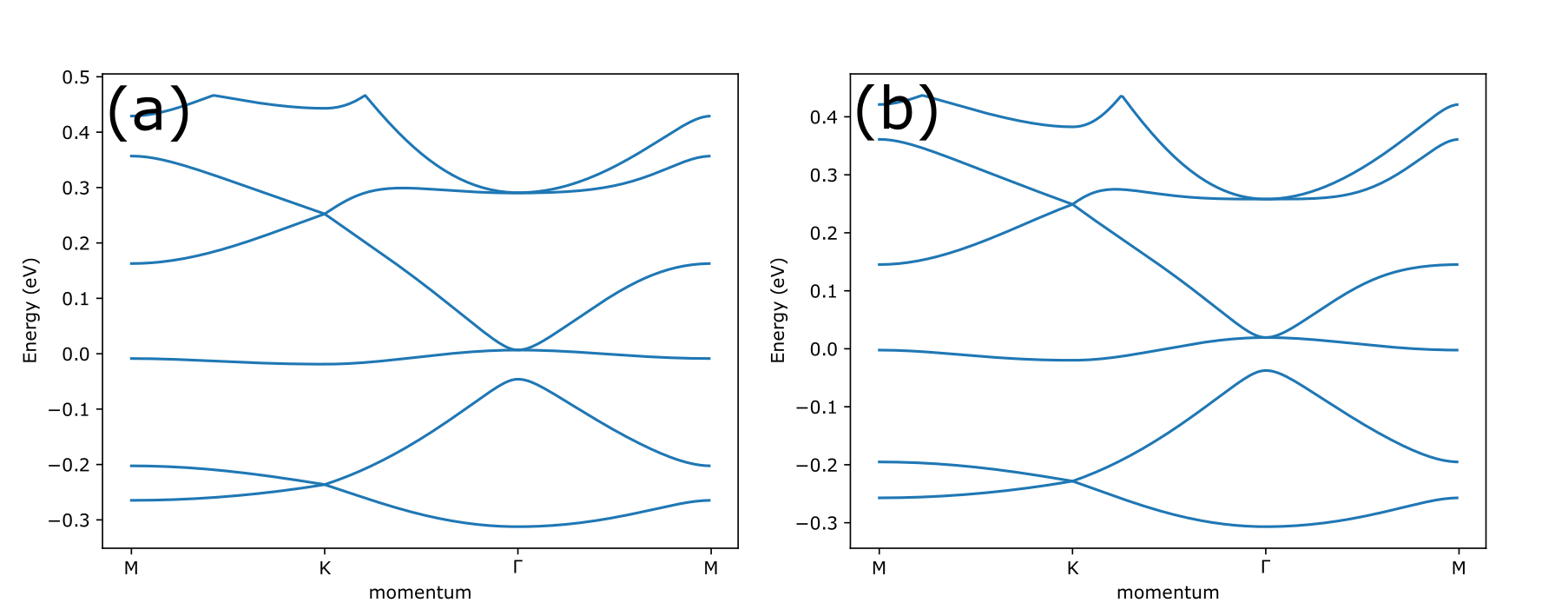}
\caption{Band structure of double ring design calculated using (a) Muffin-tin and (b) Gaussian shaped potentials. The position of CO molecules on a copper lattice is shown in Fig. \ref{fig:refsystem} (a).}
\label{fig:diffMTGA}
\end{figure}

When the spin-orbit coupling is tuned to zero, the muffin-tin and Gaussian potential can be compared. Using the same parameters as before, we indeed see the same band structures for both potentials. This is shown in Fig. \ref{fig:diffMTGA}.

\section{Rashba modified Dirac point\label{ap:Dirac}}
\begin{figure}[h]
\centering
\includegraphics[scale=0.5]{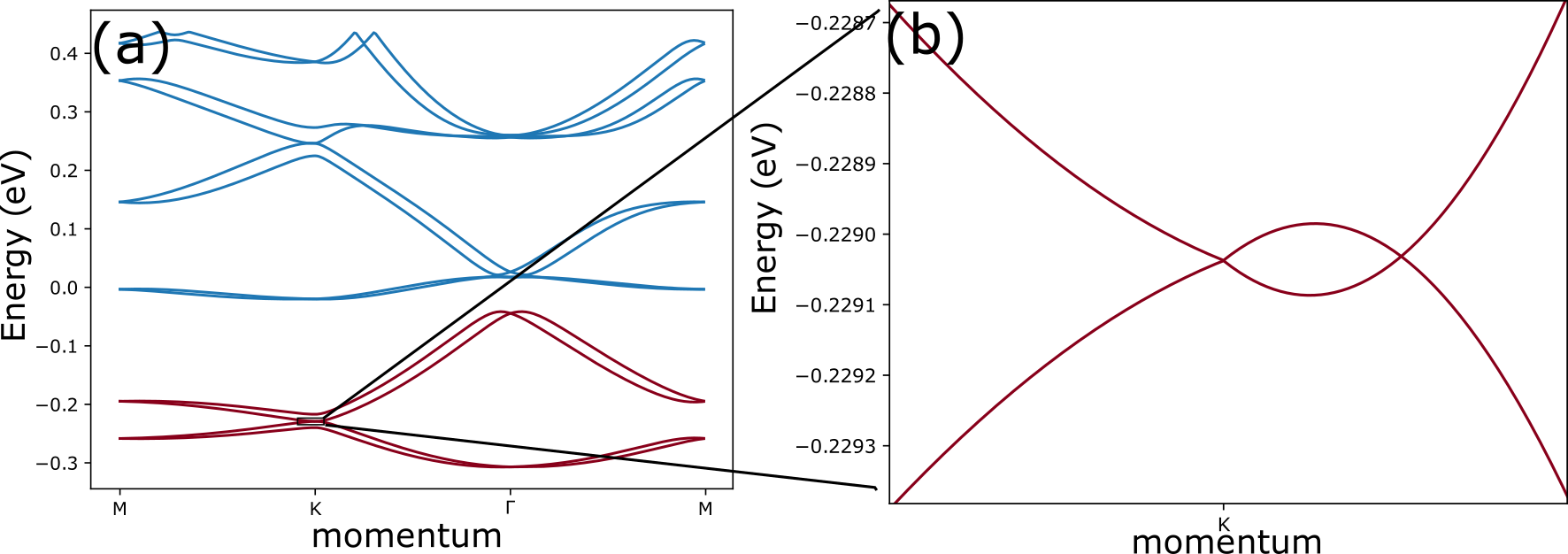}
\caption{A zoom in on the $s$ orbital Dirac cone in Fig.~\ref{fig:refsystem} (c). The location of the zoom is indicated in (a). (b) shows the zoomed in image.}
\label{fig:rashzoom}
\end{figure}

As mentioned in the main text, the Rashba coupling creates additional Dirac cones around the $s$ orbital Dirac cone in the muffin-tin method, as shown in Fig.~\ref{fig:rashzoom}. This is in agreement with tight-binding calculations for Rashba coupling in graphene \cite{van2010rashba}. 

\section{"Dirac" point between $s$ and $p$ bands\label{ap:Dirac2}}
\begin{figure}[h]
\centering
\includegraphics[scale=0.5]{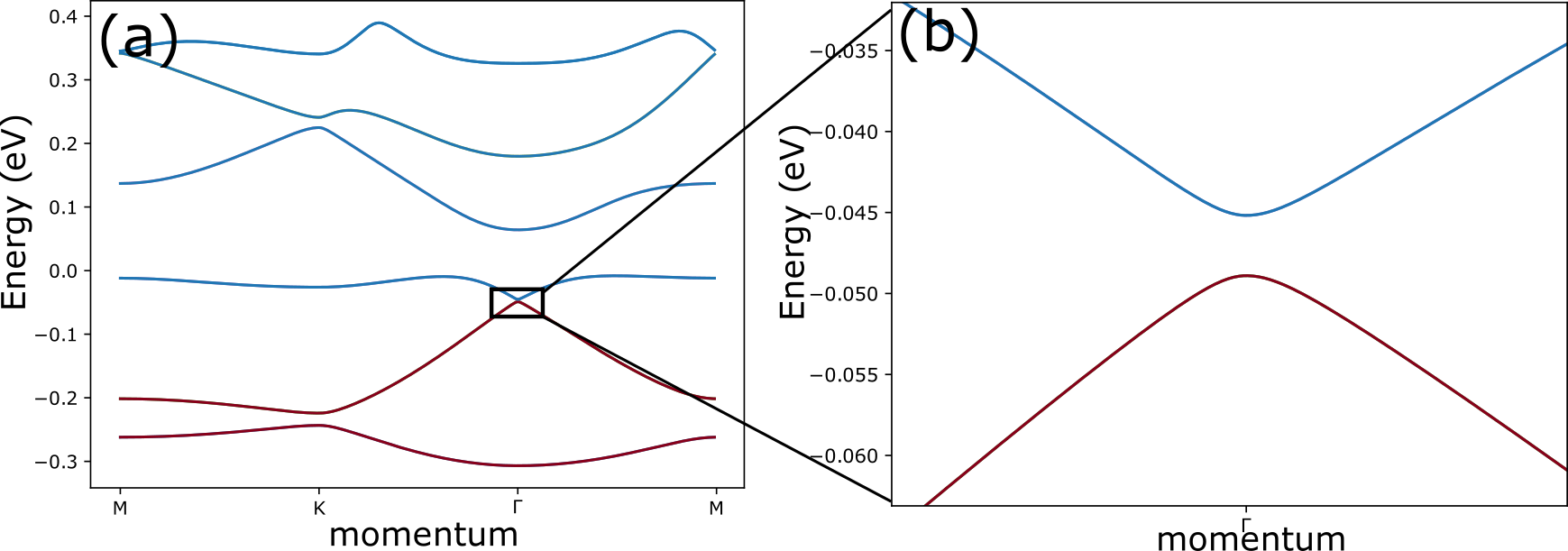}
\caption{A zoom in on the apparent Dirac cone between the $s$ and $p$ bands in Fig.~\ref{fig:refsystem} (e) of the main text. The location of the zoom is indicated in (a). (b) shows the zoomed in image.}
\label{fig:spzoom}
\end{figure}

In Fig.~\ref{fig:refsystem} (e) the $s$ and $p$ bands seem to hybridize to form a Dirac cone. However upon close inspection we see that the gap between the bands does not close, as shown in Fig.~\ref{fig:spzoom}. 

\end{document}